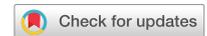

# Cardiovascular deconditioning during long-term spaceflight through multiscale modeling

Caterina Gallo [1], Luca Ridolfi[2] and Stefania Scarsoglio [1]✉

Human spaceflight has been fascinating man for centuries, representing the intangible need to explore the unknown, challenge new frontiers, advance technology, and push scientific boundaries further. A key area of importance is cardiovascular deconditioning, that is, the collection of hemodynamic changes—from blood volume shift and reduction to altered cardiac function—induced by sustained presence in microgravity. A thorough grasp of the 0G adjustment point per se is important from a physiological viewpoint and fundamental for astronauts' safety and physical capability on long spaceflights. However, hemodynamic details of cardiovascular deconditioning are incomplete, inconsistent, and poorly measured to date; thus a computational approach can be quite valuable. We present a validated 1D–0D multiscale model to study the cardiovascular response to long-term 0G spaceflight in comparison to the 1G supine reference condition. Cardiac work, oxygen consumption, and contractility indexes, as well as central mean and pulse pressures were reduced, augmenting the cardiac deconditioning scenario. Exercise tolerance of a spaceflight traveler was found to be comparable to an untrained person with a sedentary lifestyle. At the capillary–venous level significant waveform alterations were observed which can modify the regular perfusion and average nutrient supply at the cellular level. The present study suggests special attention should be paid to future long spaceflights which demand prompt physical capacity at the time of restoration of partial gravity (e.g., Moon/Mars landing). Since spaceflight deconditioning has features similar to accelerated aging understanding deconditioning mechanisms in microgravity are also relevant to the understanding of aging physiology on the Earth.

*npj Microgravity* (2020)6:27 ; https://doi.org/10.1038/s41526-020-00117-5

## INTRODUCTION

The cardiovascular system undergoes a constellation of important hemodynamic changes in microgravity—such as fluid shift of about 2 l from the legs to upper body, total blood volume reduction of around 11%, decrease in cardiac muscle mass of over 10%, and reduced exercise capability of up to 20% in the first 30 days—which lead to cardiovascular deconditioning[1–8]. Understanding 0G configuration in detail is particularly crucial to prevent and mitigate disabling effects through targeted countermeasures, and thus ensure the full health and well-being of astronauts during spaceflight missions[2]. A better comprehension of deconditioning mechanisms can answer open questions on readaptation for spaceflight beyond 1 year. It is currently unknown whether very lengthy missions lead to an amplification of reversible changes already known or the emergence of unrecognized and irreversible alterations in cardiovascular function[3]. Moreover, even if cardiovascular deconditioning is not extreme during spaceflight, it can later become hazardous at the time of reentry on the Earth or partial gravity restoration (e.g., Moon or Mars landing), when immediate demands for physical capacity are required and cannot be met. In view of future Moon or Mars missions, long-term cardiovascular deconditioning will need to be addressed to: (i) evaluate the efficacy of countermeasures and to optimize them in order to guarantee the astronauts' safety; and (ii) establish risk thresholds, especially for physical activity after landing. Additionally, as space science is an extraordinary investment multiplier (9:1 return on investment by way of spin-offs from space technology), understanding space deconditioning can have impacts on other clinical conditions and on aging research[3].

Although the overall scenario of cardiovascular deconditioning is widely accepted, the extreme difficulty of performing clinical measurements, the unsystematic data collection on older missions, the limited number of human space missions, and the heterogeneity of the missions (in terms of reference baseline posture, crew, objectives, countermeasures, duration, etc.) make the details of cardiovascular responses controversial and restricted to a modest number of cardiac parameters[9–12]. To the best of our knowledge, hemodynamic parameters related to the right heart, oxygen consumption, and exercise tolerance are barely known in microgravity. Almost no information at the vascular level regarding organ blood supply and wave propagation mechanisms is available.

Ground-based analogs—such as bed rest studies and water immersion—are not fully representative of microgravity as they introduce artificial horizontal and hydrostatic pressure gradients respectively. Parabolic flight temporarily reproduces 0G effects with each session lasting only 20–30 s[13].

Given the above difficulties, the computational approach is a recent and promising tool to study the human body's response to microgravity conditions[14–18], as it can be used to investigate variables and processes which are hard to measure, as well as predict different scenarios and optimal countermeasures. In particular, reduced order modeling seems to be the most appropriate strategy, as it is a reasonable balance between the suitable level of local detail and the overall response (see, among the most recent[19–23]). Multiscale modeling has already provided useful insights into aging[24,25], coronary pathologies[26,27], and cardiac arrhythmias[28–30].

[1]Department of Mechanical and Aerospace Engineering, Politecnico di Torino, Torino, Italy. [2]Department of Environmental, Land and Infrastructure Engineering, Politecnico di Torino, Torino, Italy. ✉email: stefania.scarsoglio@polito.it

Published in cooperation with the Biodesign Institute at Arizona State University, with the support of NASA

C. Gallo et al.

The present study aims to computationally shed light on the cardiovascular knowledge gaps previously described, following a comprehensive bibliographic investigation of more than 50 studies of cardiovascular deconditioning during microgravity. Among the available literature we only focused on the most recent long-term spaceflight data. This allowed us to extract information to characterize the steady-state cardiovascular condition for long-term (at least 5 months) spaceflight. In order to model the typical long-term response, we prioritized studies having only routine and customary countermeasures, and avoided spaceflights with specific/peculiar ad hoc countermeasures. The main cardiovascular features of the 0G adaptation point are: blood shift from lower to upper body, total blood volume reduction, reduced cardiac function and volume, increase of lower extremity venous compliance, arterial resistance variation, and baroreflex response alteration. Our goal is to quantitatively describe in detail the cardiovascular response through a comprehensive 0D–1D multi-scale model, which has been successfully tested by heart pacing and open-loop response[31]. The 0D–1D heart-arterial hemodynamics of the complete model has been previously validated to subject-specific and pathological conditions[25,28,32–34]. The physical-based model combines a 1D description of the arterial tree together with a lumped parameterization of the remaining regions, i.e., venous return, heart chambers, pulmonary circulation, and baroreceptor regulation (see the Methods for more detail and for the limitations of the present approach). In particular, we compared 1G supine and 0G spaceflight conditions in order to: (i) understand the underlying mechanisms leading to cardiovascular deconditioning, and (ii) describe the hemodynamics of zones for which clinical data are not yet feasible and accurate. The adoption of the 1G supine condition as a reference baseline has a twofold justification: (i) supine is the reference medical position defining the benchmark for physiological cardiovascular behavior in clinical practice; (ii) over recent decades, cardiovascular modeling (see, among others,[19–21,23,35]) has been derived which simulates the posture of primary medical interest and has developed a robust, data-based, and well-accepted parameter setting, which serves as a solid reference point for parameter tuning. Changes occurring in 0G condition described hereafter are always reported with respect to the (chosen) 1G supine baseline. The present study provides computational-based insights into several cardiac parameters and hemodynamic variables (in terms of time-series, average distribution throughout the body, and waveform alteration), which are currently unexplored.

## RESULTS

Results are organized showing the main cardiac parameters first. Whenever possible, to validate the modeling response we compared our results to available literature data of long-term spaceflights which employed routine countermeasures and which referred to the 1G supine condition. The cardiovascular system was divided into four areas: cerebral, cardio-thoracic, abdominal, and lower limbs. Special focus was given to the proximal-to-distal pathway of the arterial tree. Time-series, characteristic levels (i.e., mean, maximum, excursion) and waveform changes of pressures ($P$), flow rates ($Q$), and volumes ($V$), were then evaluated for representative sites of each cardiovascular region. In the following, supine 1G configuration on the Earth is depicted with blue curves, while long-term 0G spaceflight is depicted with red curves. For each configuration results are referred to a generic steady-state heartbeat, so that $t \in [0, RR]$, where $RR$ [s] is the cardiac beating period and $HR = 60/RR$ [bpm] is the heart rate.

### Cardiac parameters

The most important cardiac parameters are reported in Table 1 (see Supplementary Information for definitions). We first compared present outcomes with literature data. To this end, it should be noted that the detailed cardiovascular response in literature is quite discordant and debated. As comprehensively reported by Norsk[12], for different hemodynamic parameters—such as heart rate, arterial pressure, cardiac output, stroke volume—there are conflicting results, which depend on the level of adopted countermeasures, the mission duration, and, most importantly, the baseline (upright/sitting/supine) reference 1G condition. In view of the above considerations very few assessed metrics were available for full comparison and, among these, we selected the most recent studies involving long-term spaceflights (around 5 months). Thus, collected measurements are the most reliable results which we adopted as reference values for the present study.

A first validation comes from the left ventricular contractility parameters: they all reduced (−18.32% for $SV$, −9.30% for $V_{lved}$, −9.93% for $EF$) except for a slight increase of $V_{lves}$ +4.75%, and are in good agreement with spaceflight data[4,5,36–38]. A reduction of cardiac contractility is also in general accordance with the cardiac atrophy observed during bed rest[39–41]. $CO$ drop (−7.58%) is less evident than the $HR$ increase in the spaceflight configuration (+13%, see the Methods section). The $CO$ decrease is in qualitative agreement with the variations observed with respect to 1G supine condition[4,5,36–38]. As previously mentioned, $CO$ (together with $SV$) is one of the cardiac parameters whose variation in 0G greatly depends on the chosen 1G baseline configuration. As observed by ref.[12,42], in 0G $CO$ increased with respect to the sitting 1G position. Compared to the supine 1G position, $CO$ was usually observed to decrease[4,5,36–38], although mild increases were also reported[42]. Mean central arterial pressure ($MAP$) was also tested for validation: $MAP$ decreased by 9.92% in our model (as does brachial arterial pressure), falling within the range of measured data during long-term spaceflights[4,6,43,44]. Mild reduction of $MAP$ with respect to seated preflight values was also observed in long-term spaceflight using a countermeasure protocol[1,12]. The green rows of Table 1 show that there is satisfactory agreement between currently available in vivo and our in silico outcomes. This confirms the correct 0G configuration setting and the model's accuracy. On the other hand, the scarcity of previously measured parameters and the good predictive level of the model justifies its use to obtain cardiac estimates not previously measured directly. It should be recalled that these cardiac variables were not used to tune the model parameters, but to a posteriori verify the modeling response once the parameter setting was completed. The comparison with the reported literature represents the only validation (with data actually measured and related to the same simulated configuration) currently available in 0G conditions.

Other cardiac parameters were also modeled. A substantial decrease of the work done by the heart (−19.29% for $SW/min$) and a more limited reduction of oxygen consumption (−1.91% for $RPP$, −8.38% for $TTI/min$) was found. The slight decrease of myocardial oxygen consumption indexes represents an additional modeling validation, in agreement with the maximal oxygen uptake reduction observed in long-term spaceflight[45]. Central venous pressure ($CVP$), estimated as the mean value of the superior and inferior vena cava pressure signals averaged over the beat, decreased by 5.08%. This result is in qualitative agreement with the $CVP$ reduction observed in early spaceflights[12,46,47] and parabolic flights[48,49], though it should be kept in mind that the time scales and physical mechanisms inducing $CVP$ variations found in literature are not fully comparable to the present study. Central aortic pulse pressure, $PP_{AA}$, decreased (−22.30%) due to a higher reduction of central aortic systolic pressure (−13.39%, with respect to the diastolic pressure decrement −6.90%). In agreement with the pulse pressure behavior, central aortic augmentation index, $AI_{AA}$, also declined (−41.18%).

The observed variations of cardiac work, oxygen consumption, and contractility indexes, as well as central mean and pulse

Published in cooperation with the Biodesign Institute at Arizona State University, with the support of NASA

**Table 1.** Modeled cardiac parameters for the supine 1G configuration and during long-term 0G spaceflight (green color is adopted when spaceflight data are available, yellow otherwise).

| Variable | Supine 1G | Spaceflight 0G | % Variation | Literature data |
|---|---|---|---|---|
| $V_{lves}$ [ml] | 48.43 | 50.73 | + 4.75% | – |
| $V_{lved}$ [ml] | 123.96 | 112.43 | −9.30% | $[-5\%, -13\%]^{4, 5, 36-38}$ |
| $SV$ [ml] | 75.54 | 61.70 | −18.32% | $[-14\%, -23\%]^{4, 36-38}$ |
| $EF$ [%] | 60.93 | 54.88 | −9.93% | $[-5\%, -11\%]^{4, 5, 36-38}$ |
| $CO$ [l/min] | 5.67 | 5.24 | −7.58% | $[-11\%, -18\%]^{4, 36-38}$ |
| $SW/min$ [J/min] | 80.01 | 64.58 | −19.29% | – |
| $CVP$ [mmHg] | 6.49 | 6.16 | −5.08% | – |
| $P_{AA,syst}$ [mmHg] | 121.03 | 104.82 | −13.39% | – |
| $P_{AA,dias}$ [mmHg] | 69.99 | 65.16 | −6.90% | – |
| $MAP$ [mmHg] | 87.01 | 78.38 | −9.92% | $[-2\%, -10\%]^{4, 6, 43, 44}$ |
| $TTI/min_{[mmHg\ s/min]}$ | 2590 | 2373 | −8.38% | – |
| $RPP$ [mmHg/min] | 9078 | 8905 | −1.91% | – |
| $AI_{AA}$ | 0.17 | 0.10 | −41.18% | – |
| $PP_{AA}$ [mmHg] | 51.04 | 39.66 | −22.30% | – |

$V_{lves}$ end-systolic left ventricular volume, $V_{lved}$ end-diastolic left ventricular volume, $SV$ stroke volume, $EF$ ejection fraction, $CO$ cardiac output, $SW/min$ stroke work per minute, $CVP$ central venous pressure, $P_{AA,syst}$ and $P_{AA,dias}$ systolic and diastolic ascending aortic pressures, $MAP$ mean arterial pressure, $TTI/min$ tension time index per minute, $RPP$ rate pressure product, $AI_{AA}$ and $PP_{AA}$ augmentation index and pulse pressure for the ascending aortic district.

pressures signal an overall scenario of cardiac deconditioning. The cardiovascular system reaches a condition commonly seen in a sedentary lifestyle. This is achieved initially by direct fluid dynamics changes and is then maintained through continued presence in the less demanding 0G environment.

**Pressures, flow rates and volumes**
Pressure, $P(t)$, flow rate, $Q(t)$, and volume, $V(t)$, time-series were analyzed together with the relative variations of the beat-averaged values ($\bar{P}$, $\bar{Q}$, and $\bar{V}$) during spaceflight with respect to the supine 1G configuration on the Earth. In addition to mean variation, relative variation of pulse pressure, $PP$ (defined, in the general case, as the difference between maximum and minimum pressure values), and maximum pressure values, $P_{max}$, were evaluated. In addition to being general indicators of health and being mechanical properties of the cardiovascular system, these latter quantities estimate the vertical stretching of the pressure signals. Figure 1 shows representative results for the four macro-regions considered, while $P$ and $Q$ variations at other significant sites are listed in Tables 2 and 3.

Both pressure and flow rate mean levels decreased in the spaceflight configuration. Pressure variations were quite homogeneous throughout the different cardiovascular regions, ranging between −8 and −12% (apart from the cardiac region and venous return, where variations are lower, between −2 and −6%). Flow rate changes were more heterogeneous. In particular, variations of the mean values decreased from the upper to the lower body, from −16 to −18% in the cerebral region to −2 to −6% in the lower limbs (apart from the anterior tibial artery, −11.94%). Pressure maximum variations are more heterogeneous (around −5 to −15%) than mean levels, and they are both smaller in the venous return than in the arterial circulation. $PP$ variations remarkably showed high variability (ranging from +4.82 to −56.00%), which is not directly attributable to the location, type (venous/arterial), and size of the region. An example is represented by the superior and inferior venae cavae (see Table 3), both veins ending in the right atrium and similar in size, but showing completely different $PP$ variations (+4.82 and −32.97%).

Volume time-series, $V(t)$, and relative variations of the average values per beat ($\bar{V}$) during spaceflight with respect to the supine 1G condition are reported in Fig. 2 for the 0D compartments. Mean pressure and flow rate values are strictly related to blood volume variations induced by the prolonged exposure to microgravity, since $V = V_0 + CP$, where $C$ and $V_0$ are the compliance and unstressed volume of each district, respectively. In the lower limbs and abdominal region below VIP (VIP, see the Methods for more details), mean volume variations were widely negative. In the upper body, volume reduction was smaller, resulting in volume increase in the upper body and pulmonary regions. The different upper-to-lower body volume repartition primarily resulted from the blood shift, which in the upper body partially counterbalances the effects of volume reduction. The further cardiac volume and contractility reduction mitigated the fluid shift effects in the cardiac region. As a result, heart chamber volumes were reduced more than the thoracic surroundings.

The time-series of the hemodynamic variables shown in Figs. 1 and 2 demonstrate the whole signal differences between 1G supine and 0G conditions. Signals were not merely translated or scaled but, also due to the different $HR$, their waveforms changed. This aspect is discussed in more detail in the next section.

**Hemodynamic signals: waveform alteration**
We focused on the waveform alteration of pressure and flow rate signals in terms of the normalized signal difference, $NSD$, which is a dimensionless measure of signaling similarity (ranging in the interval [0.2]) between the 1G supine and 0G spaceflight configurations (see the Methods section for more details). Representative results of each cardiovascular region are shown in Fig. 3, while $NSD$ values for both $P'$ and $Q'$ at different sites can be found in Table 4.

It is immediately apparent that the normalized signals $P'$ e $Q'$ in 0G (red curves) in all cardiovascular sites were delayed with respect to the 1G supine condition (blue curves), as a result of the horizontal stretching due to the faster $RR$ beating rate in 0G. However, the delay was not merely due to $HR$ variations, as it was not uniformly found in all the regions, but varied without showing

Published in cooperation with the Biodesign Institute at Arizona State University, with the support of NASA

npj Microgravity (2020) 27

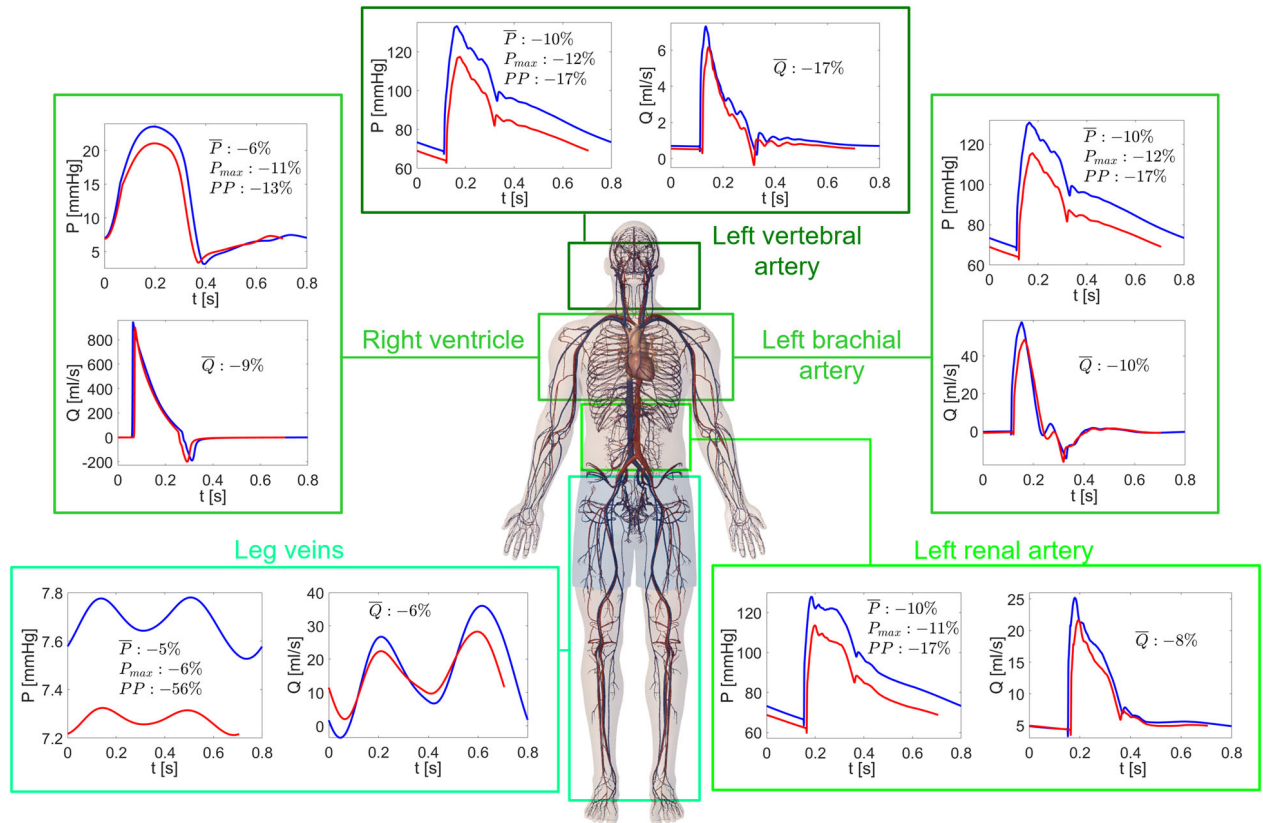

**Fig. 1 Pressures and flow rates throughout the body in 1G supine and 0G conditions.** Time-series of the pressure, $P(t)$, and flow rate, $Q(t)$, at different sites: left vertebral artery (cerebral), right ventricle and left brachial artery (cardio-thoracic region), left renal artery (abdominal region), leg veins (lower limbs). Blue: supine 1G configuration on the Earth; red: 0G spaceflight configuration. Mean pressure ($\bar{P}$), mean flow rate ($\bar{Q}$), together with maximum pressure ($P_{max}$), and pulse pressure ($PP$) relative variations between 1G and 0G configurations are reported in each panel.

a significant trend from upper to lower regions. This means that waveform alteration acts differently throughout the body and as well as at the same site. For instance, by comparing 0G and 1G, note the different shape and phase of the $Q'$ signals in the inferior vena cava, and the different oscillation amplitude for the carotid $P'$ and $Q'$ signals. *NSD*, which was introduced to quantify signal stretching besides pure horizontal translation, was generally not negligible and often beyond 0.5. Flow rates were more affected by shape variations than pressures, and proximal regions were less prone to waveform variations than distal regions (see Table 4).

Arterial proximal-to-distal tree

Special attention was given to the arterial proximal-to-distal tree, focusing on the arterial 1D system from the ascending aorta to the anterior tibial artery. The scenario emerging from Fig. 4 (and detailed in Tables 2 to 4) is multi-faceted and not always ascribable to the macro-regions crossed by the pathway. Mean flow rate variations were close to −9% in the ascending aorta, increasing to approximately −6% in the last part of the abdominal aorta, and then decreasing again towards the anterior tibial artery to around −12%. Mean pressure variations were instead quite constant, in the range of −10%, along the pathway, similarly to what happens for the $P_{max}$ (−11 to −13%). The overall proximal-to-distal trend of $PP$ variations was a slight decrease, while *NSD* variations increased remarkably. In particular, *NSD* at least doubled for $P'$ and $Q'$, and $Q'$ variations were also larger than for $P'$. What was previously reported for the $P'$, $Q'$, and *NSD* behavior from proximal-to distal areas is also true here, where only arterial sites were considered.

## DISCUSSION

These results depict a complex picture where changes imposed by the 0G configuration created remarkably heterogenous stresses on the hemodynamic variables of the cardiovascular system. Volume reduction and baroreflex responses acted on global levels of pressure, volume, and flow rate, while blood shift, resistance and compliance variations altered the local mean levels of the hemodynamic variables throughout the body. The most important alteration in volume was the combination of blood shift and reduction: in the upper body the incoming blood volume partially compensated for the volume lost, resulting in either small reductions or positive variations (in particular, at the pulmonary and cerebral levels). In the lower body two contributions added up, leading to a more evident volume decrease. However, the upper body blood volume overload did not translate into pressure and flow rate increases due to the complex interplay between resistance and compliance variations within each region. Moreover, it should be kept in mind that (negative) variations were referred to the 1G supine condition, while in the cerebral circulation positive variations are expected if referred to the 1G standing position, as suggested by ref. [50]. The increased venous compliance observed in the legs confined lower body mean pressure variations to a range which is similar to that of the upper body, that is −8 to −12% (see Fig. 1 and Table 2). The smallest variations (−5%) occurred in the capillary–venous districts of the legs. Cardiac compliance increased accounting for the reduced contractility, and this in turn limited pressure variations at the central level (−1 to −8%). If arterial resistance had not changed, following homogeneous pressure variations we would have expected homogeneous flow

Published in cooperation with the Biodesign Institute at Arizona State University, with the support of NASA

**Table 2.** Beat-averaged levels of pressure $\bar{P} = \int_{RR} P(t)dt/RR$ [mmHg] and flow rate $\bar{Q} = \int_{RR} Q(t)dt/RR$ [ml/s] at different significant sites in supine 1G and 0G conditions, with relative variations [%].

| Cardiovascular region | Pressure $\bar{P}$ [mmHg] | | | Flow rate $\bar{Q}$ [ml/s] | | |
|---|---|---|---|---|---|---|
| | 1G | 0G | Variation [%] | 1G | 0G | Variation [%] |
| Arterial tree pathway (1D) | | | | | | |
| Ascending aorta | 92.61 | 83.19 | −10.17% | 87.85 | 80.01 | −8.92% |
| Thoracic aorta | 92.38 | 82.99 | −10.16% | 67.07 | 62.20 | −7.26% |
| Abdominal aorta A | 90.29 | 81.12 | −10.16% | 46.95 | 44.04 | −6.20% |
| Abdominal aorta E | 90.76 | 81.45 | −10.26% | 17.01 | 16.02 | −5.82% |
| External iliac | 90.48 | 80.93 | −10.55% | 6.31 | 5.86 | −7.13% |
| Femoral | 89.47 | 80.20 | −10.36% | 3.24 | 2.94 | −9.26% |
| Cerebral region (0D–1D) | | | | | | |
| HA veins | 6.94 | 6.54 | −5.76% | 20.82 | 17.86 | −14.22% |
| Left vertebral | 92.47 | 83.13 | −10.10% | 1.60 | 1.33 | −16.88% |
| Left vertebral arterioles | 70.60 | 64.86 | −8.13% | 1.61 | 1.34 | −16.77% |
| Left internal carotid | 92.27 | 83.03 | −10.01% | 2.16 | 1.76 | −18.52% |
| Left internal carotid arterioles | 80.40 | 73.32 | −8.81% | 2.18 | 1.78 | −18.35% |
| Left external carotid | 92.28 | 83.05 | −10.00% | 2.34 | 1.90 | −18.80% |
| Left external carotid arterioles | 80.37 | 73.32 | −8.77% | 2.35 | 1.93 | −17.87% |
| Cardio-thoracic region (0D–1D) | | | | | | |
| LA and MV | 7.51 | 7.39 | −1.60% | 87.91 | 80.11 | −8.87% |
| LV and AV | 43.20 | 39.59 | −8.36% | 87.85 | 80.01 | −8.92% |
| RA and TV | 6.47 | 6.14 | −5.10% | 87.90 | 80.15 | −8.82% |
| RV and PV | 11.64 | 10.95 | −5.93% | 87.84 | 80.04 | −8.88% |
| Pulmonary arteries | 15.42 | 14.59 | −5.38% | 87.83 | 80.02 | −8.89% |
| Pulmonary veins | 8.39 | 8.19 | −2.38% | 87.86 | 80.05 | −8.89% |
| SVC | 6.48 | 6.15 | −5.09% | 20.80 | 17.85 | −14.18% |
| IVC | 6.50 | 6.17 | −5.08% | 67.16 | 62.35 | −7.16% |
| Intercostals | 92.64 | 83.22 | −10.17% | 9.93 | 8.85 | −10.88% |
| Intercostals arterioles | 85.39 | 76.74 | −10.13% | 9.89 | 8.85 | −10.52% |
| Left brachial | 92.33 | 82.97 | −10.14% | 4.29 | 3.85 | −10.27% |
| Abdominal region and organs (0D–1D) | | | | | | |
| Gastric | 89.36 | 80.44 | −9.98% | 4.92 | 4.41 | −10.37% |
| Gastric arterioles | 48.88 | 44.05 | −9.88% | 4.92 | 4.42 | −10.16% |
| Celiac A | 91.57 | 82.26 | −10.17% | 10.17 | 9.19 | −9.63% |
| Left renal | 90.43 | 81.23 | −10.17% | 8.50 | 7.81 | −8.12% |
| U left renal arterioles | 66.18 | 58.90 | −11.00% | 4.19 | 3.68 | −12.17% |
| L left renal arterioles | 66.18 | 58.90 | −11.00% | 4.30 | 4.13 | −3.95% |
| U ABD venules | 12.94 | 11.90 | −8.04% | 26.35 | 23.44 | −11.04% |
| L ABD venules | 12.98 | 12.38 | −4.62% | 23.64 | 22.68 | −4.06% |
| Lower limbs region (0D–1D) | | | | | | |
| Inner iliac | 90.47 | 81.17 | −10.28% | 2.20 | 2.15 | −2.27% |
| Inner iliac arterioles | 78.65 | 69.59 | −11.52% | 2.20 | 2.15 | −2.27% |
| Deep femoral | 89.44 | 80.17 | −10.36% | 3.08 | 2.93 | −4.87% |
| Deep femoral arterioles | 68.12 | 59.74 | −12.30% | 3.07 | 2.93 | −4.56% |
| Anterior tibial | 84.44 | 75.61 | −10.46% | 2.01 | 1.77 | −11.94% |
| Anterior tibial arterioles | 38.20 | 34.16 | −10.58% | 2.01 | 1.78 | −11.44% |
| Posterior tibial | 84.90 | 76.00 | −10.48% | 1.22 | 1.15 | −5.74% |
| Posterior tibial arterioles | 64.36 | 56.33 | −12.48% | 1.23 | 1.16 | −5.69% |
| Legs capillaries | 27.13 | 25.63 | −5.53% | 17.02 | 16.04 | −5.76% |
| Legs veins | 7.68 | 7.28 | −5.21% | 17.03 | 16.04 | −5.81% |

**Table 3.** Maximum pressure, $P_{max}$, and pulse pressure, $PP$, values [mmHg] at different significant sites in supine 1G and 0G conditions, with relative variations [%].

| Cardiovascular region | $P_{max}$ [mmHg] | | | $PP$ [mmHg] | | |
|---|---|---|---|---|---|---|
| | 1G | 0G | Variation [%] | 1G | 0G | Variation [%] |
| Arterial tree pathway (1D) | | | | | | |
| Ascending aorta | 121.03 | 104.82 | −13.39% | 51.04 | 39.66 | −22.30% |
| Thoracic aorta | 129.83 | 114.49 | −11.82% | 61.77 | 51.04 | −17.37% |
| Abdominal aorta A | 123.92 | 109.98 | −11.25% | 58.75 | 48.83 | −16.89% |
| Abdominal aorta E | 135.38 | 119.27 | −11.90% | 72.78 | 60.26 | −17.20% |
| External iliac | 135.07 | 119.46 | −11.56% | 74.55 | 62.15 | −16.63% |
| Femoral | 135.63 | 119.75 | −11.71% | 76.97 | 63.71 | −17.23% |
| Cerebral region (0D–1D) | | | | | | |
| HA veins | 7.33 | 6.84 | −6.68% | 0.72 | 0.64 | −11.11% |
| Left vertebral | 133.17 | 117.39 | −11.85% | 65.80 | 54.55 | −17.10% |
| Left vertebral arterioles | 82.67 | 74.08 | −10.39% | 25.66 | 19.77 | −22.95% |
| Left internal carotid | 140.36 | 124.19 | −11.52% | 74.99 | 62.99 | −16.00% |
| Left internal carotid arterioles | 101.30 | 90.77 | −10.40% | 39.52 | 32.47 | −17.84% |
| Left external carotid | 141.51 | 125.13 | −11.58% | 76.19 | 63.99 | −16.01% |
| Left external carotid arterioles | 100.25 | 89.30 | −10.92% | 38.03 | 30.53 | −19.72% |
| Cardio-thoracic region (0D–1D) | | | | | | |
| LA and MV | 8.67 | 8.54 | −1.50% | 2.49 | 2.27 | −8.84% |
| LV and AV | 121.27 | 105.10 | −13.33% | 118.26 | 101.38 | −14.27% |
| RA and TV | 7.93 | 7.42 | −6.43% | 3.12 | 3.12 | − |
| RV and PV | 23.56 | 21.08 | −10.53% | 20.40 | 17.73 | −13.09% |
| Pulmonary arteries | 23.35 | 20.92 | −10.41% | 13.49 | 10.93 | −18.98% |
| Pulmonary veins | 9.10 | 8.68 | −4.62% | 1.21 | 0.83 | −31.40% |
| SVC | 9.09 | 8.59 | −5.50% | 4.77 | 5.00 | +4.82% |
| IVC | 7.91 | 7.14 | −9.73% | 2.73 | 1.83 | −32.97% |
| Intercostals | 141.78 | 124.59 | −12.12% | 74.57 | 61.88 | −17.02% |
| Intercostals arterioles | 115.90 | 103.48 | −10.72% | 51.63 | 43.67 | −15.42% |
| Left brachial | 130.85 | 115.57 | −11.68% | 63.59 | 52.85 | −16.89% |
| Abdominal region and organs (0D–1D) | | | | | | |
| Gastric | 130.97 | 116.02 | −11.41% | 67.09 | 56.02 | −16.50% |
| Gastric arterioles | 56.93 | 50.23 | −11.77% | 16.94 | 13.02 | −23.14% |
| Celiac A | 135.66 | 119.61 | −11.83% | 70.55 | 58.49 | −17.09% |
| Left renal | 127.89 | 113.58 | −11.19% | 64.44 | 53.80 | −6.51% |
| Left renal arterioles (U and L) | 78.37 | 68.11 | −13.09% | 24.92 | 18.91 | −24.12% |
| U ABD venules | 13.64 | 12.39 | −9.16% | 1.60 | 1.15 | −28.12% |
| L ABD venules | 13.81 | 13.03 | −5.65% | 1.74 | 1.38 | −20.69% |
| Lower limbs region (0D–1D) | | | | | | |
| Inner iliac | 137.60 | 121.48 | −11.72% | 76.79 | 63.97 | −16.69% |
| Inner iliac arterioles | 105.54 | 90.17 | −14.56% | 46.21 | 35.46 | −23.26% |
| Deep femoral | 136.30 | 120.03 | −11.94% | 77.21 | 63.70 | −17.50% |
| Deep femoral arterioles | 86.66 | 73.92 | −14.70% | 33.69 | 25.66 | −23.83% |
| Anterior tibial | 144.32 | 124.92 | −13.44% | 91.07 | 73.91 | −18.84% |
| Anterior tibial arterioles | 45.93 | 40.33 | −12.19% | 14.13 | 10.91 | −22.79% |
| Posterior tibial | 144.91 | 125.36 | −13.49% | 91.46 | 74.11 | −18.97% |
| Posterior tibial arterioles | 85.48 | 72.87 | −14.75% | 35.96 | 27.65 | −23.11% |
| Legs capillaries | 31.59 | 29.23 | −7.47% | 8.30 | 6.71 | −19.16% |
| Legs veins | 7.78 | 7.32 | −5.91% | 0.25 | 0.11 | −56.00% |

rate variations throughout the body, around −8 to −12%. Instead, all lower body resistances reduced, while carotid/vertebral resistances increased. Therefore, flow rates towards upper body and cerebral areas were reduced (−17 to −19%), while lower body flow rate was on average boosted (−2 to −5%, with the anterior tibial artery as only exception). The cardio-thoracic region experienced flow rate reductions in between lower and upper body values, namely −8 to −10%.

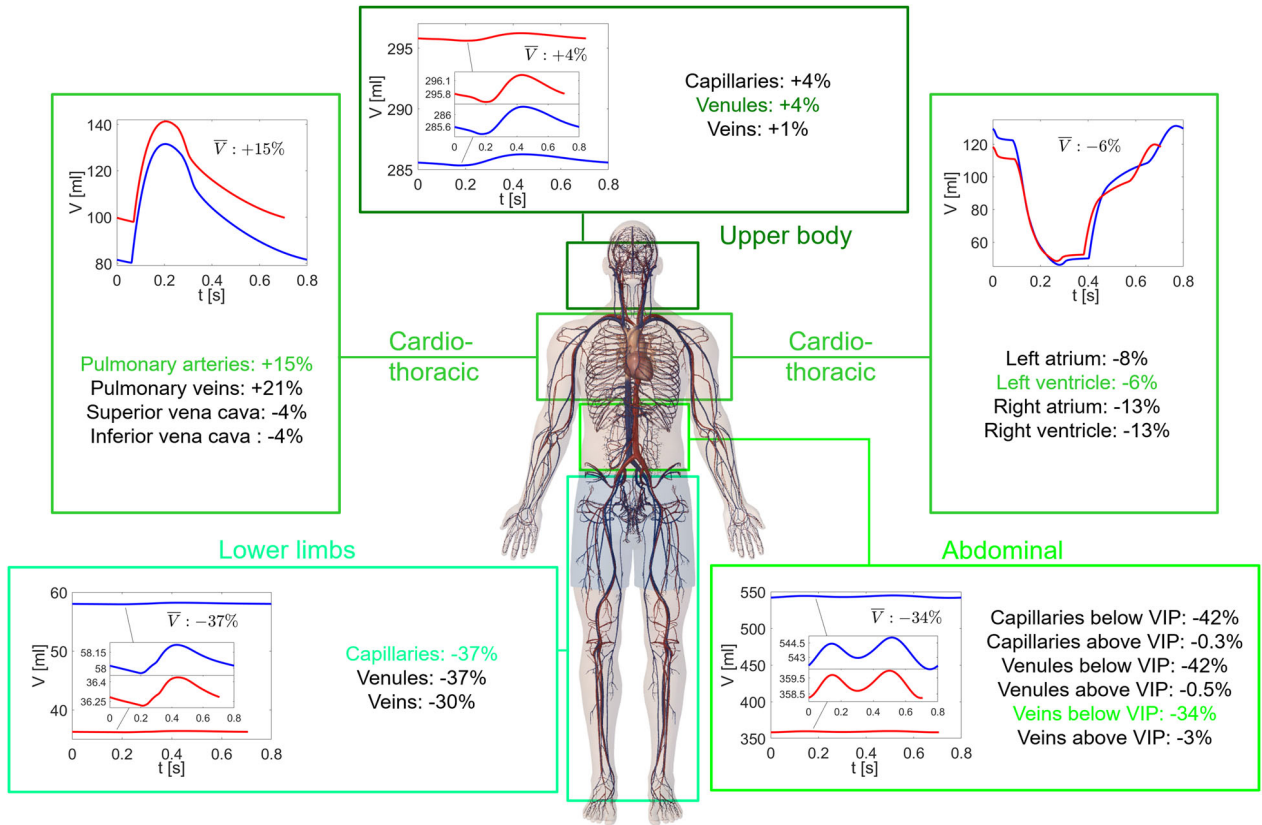

**Fig. 2  Volumes throughout the body in 1G supine and 0G conditions.** Time-series of the volume, $V(t)$, at different 0D sites: venules (upper body), left ventricle and pulmonary artery (cardio-thoracic region), veins below VIP (abdominal region), capillaries (lower limbs). Blue: supine 1G configuration on the Earth; red: 0G spaceflight configuration. Mean volume ($\bar{V}$) relative variations between 1G and 0G configurations are reported for each region (colored values refer to the corresponding time-series shown).

Mean flow rate variations decreased from ascending aorta to the inferior abdominal aorta, consistent with what happened from upper towards lower body, i.e., due to resistance changes flow rate was more hampered in the upper thoracic region with respect to the lower abdomen. Note also that ascending aorta variations were damped by the cardiac volume and contractility reduction. Mean flow rate variations then increased again towards the inner iliac-femoral-posterior tibial path, for two possible reasons: (i) conduction 1D arteries (e.g., femoral and external iliac arteries) are less influenced by resistance variations than their bifurcating branches (e.g., deep femoral and inner iliac arteries); (ii) on the Earth, resistance of the posterior tibial artery is higher than the anterior tibial. Equal relative reduction of resistance (−10%) for both arteries led to an absolute resistance reduction which is more important for the posterior tibial artery. In the latter, flow rate was thus more enhanced (i.e., lower mean $Q$ variations) than in the anterior tibial artery (higher mean $Q$ variations).

The hemodynamic and cardiac parameter variability reveals a pattern which, overall, resembles the spaceflight configuration of a relaxed, sedentary lifestyle. This is directly due to both fluid dynamics effects as well as reduced physical activity and lower cardiac work to counter the force of gravity[3]. Cardiac work, oxygen consumption, and contractility indexes were reduced, as were central mean and pulse pressures. Almost all the hemodynamic and cardiac parameters resulted from variables pertaining to the thoracic-cardiac region, for which the above considerations about $P$ and $Q$ mean values still hold. The hemodynamic mechanisms mainly responsible for the observed cardiac deconditioning are the global volume reduction together with the decrease of cardiac volume and contractility. The slight increase of $V_{lves}$ can be interpreted as a possible sign of a reduced ability of the ventricle to empty and ventricular overload due to fluid shift (pulmonary artery and vein volumes increased on average by 14.61% and 21.39%, respectively).

Maximum and mean pressure value variations were smaller in the venous than in the arterial compartments (see Tables 2 and 3). This behavior is related to the higher compliance of the venous regions, which was further increased in the lower limbs during spaceflight. Venous vessels were thus more able to absorb and reduce pressure changes than their arterial counterparts. In general, PP were higher than mean pressure and $P_{max}$ changes: apart from one exception (+4.82%), pressure signals tended to reduce their ranges to a greater extent than their means. In general signals were less pulsatile (between −10 and −30% from 1G supine to 0G) and, in each region, the greatest PP variations (e.g., >20%) usually occurred in the distal areas, where PP on the Earth is already limited and among the lowest in that region: arterioles, venules and veins were thus the most affected areas. However, the contrary was not in general true: the smallest PP values on the Earth (e.g., right atrium and superior vena cava) did not necessarily imply the largest PP variations during spaceflight.

PP and NSD variations showed great variability throughout the body, meaning that signals were stretched differently in vertical and horizontal directions depending on the site. PP and NSD alterations were inverted along the 1D arterial tree (see Fig. 4 and Tables 3, 4). At one extreme (ascending aorta) PP variation was high and signal was more squeezed than in distal regions, but NSD variations were limited. At the other extreme (anterior tibial artery), pressure signal was stretched less but experienced a remarkable waveform change in terms of NSD.

Published in cooperation with the Biodesign Institute at Arizona State University, with the support of NASA

npj Microgravity (2020) 27

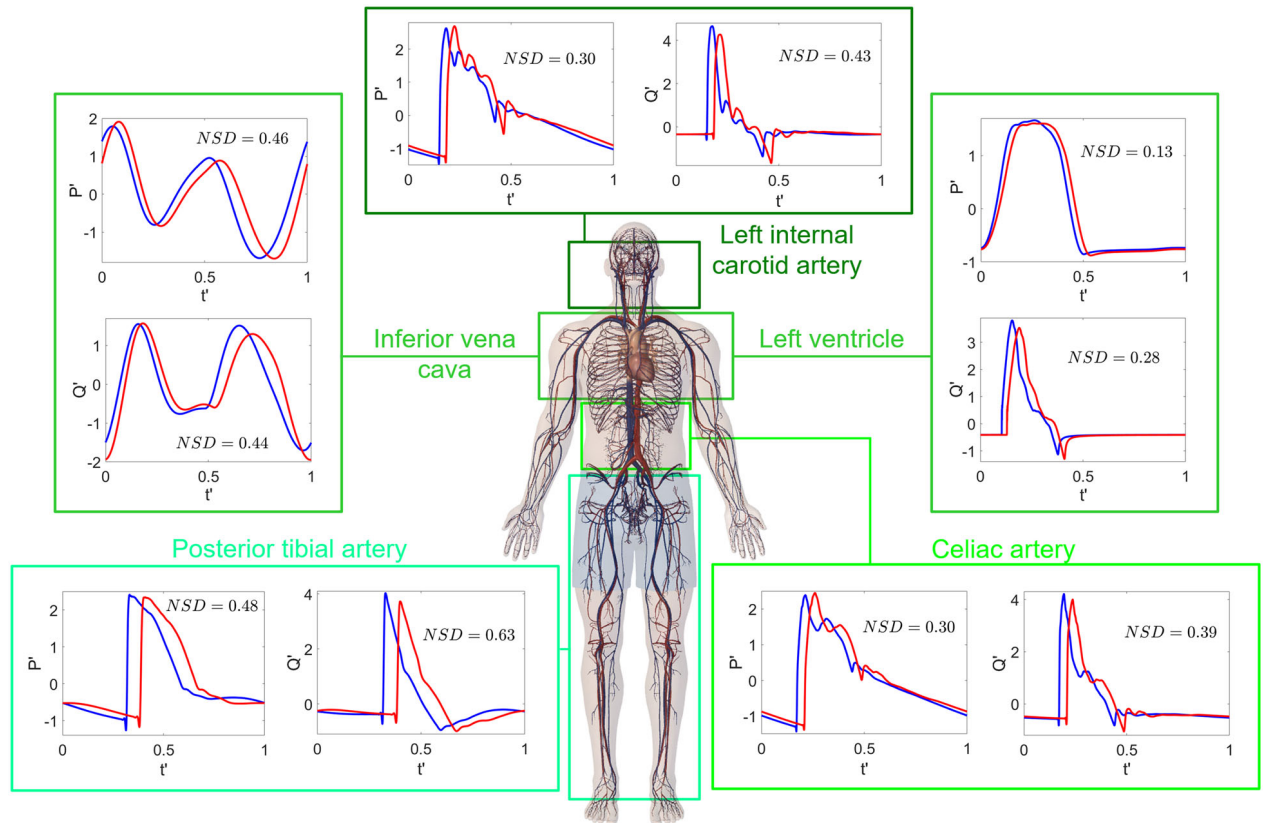

**Fig. 3  Pressure and flow rate waveform alterations.** Normalized time-series $P'(t')$ and $Q'(t')$ at different sites: left internal carotid artery (cerebral), left ventricle and inferior vena cava (cardio-thoracic region), celiac artery (abdominal region), posterior tibial artery (lower limbs). Blue: supine 1G configuration on the Earth; red: 0G spaceflight configuration. $NSD$ (normalized signal difference) values for $Q'$ and $P'$ are reported in each panel.

A possible interpretation for the $NSD$ increase towards the distal regions is that 0G-induced waveform alterations mainly happen peripherally, where the traveling wave is already modified in 1G by vessel tapering and non-linearities[51], and the great majority of reflections occur[52]. In addition, since distal regions are most closely influenced by arterial resistance and venous compliance variations, more pronounced alterations can be expected there. Thus, in the proximal regions the interaction of many backward waves arriving at different times makes them dampen and partially elide each other. As a result, the total wave was shaped mainly by the forward contribution, and waveform variation induced by the 0G configuration was less evident ($NSD < 0.2$). When instead, in the peripheral zones, forward and backward waves are comparable due to the proximity to reflection sources (such as bifurcations, discontinuities, or arteriolar interface), $NSD$ was up to or beyond 0.5. The backward wave being most influenced during the 0G configuration explains the greater $NSD$ variations for $Q$ than $P$. In fact, the flow rate backward wave is upside down with respect to the corresponding forward wave, while the pressure backward wave is reflected upright. Thus, the same backward wave variation (in absolute amplitude) has a stronger impact on the total flow rate wave rather than on pressure.

The $NSD$ increase observed in the peripheral regions is likely due to reflection mechanisms within the traveling wave combined with the $HR$ increase which, during spaceflight, exacerbated the waveform alteration. Moreover, due to the $HR$ increase in 0G the systolic interval was longer (with respect to the $RR$ period) than in 1G. Even if the imposed $HR$ increase was not worrisome ($+13\%$), diastole was shortened and diastolic function (e.g., ventricular filling) could be partly affected. As previously noted, waveform variations induced by reflection mechanisms were not absorbed downstream in the arterial tree, leading to evident signal variations (e.g., $NSD$ values around 0.4–0.5) in arteriolar–capillary districts, leg veins, and venae cavae. It is reasonable to expect that these regions downstream of the arterial system do nothing but undergo and maintain the changes introduced upstream. The normal wave pattern was therefore modified and all the physiological phenomena at the capillary–venous level—such as regular perfusion, mean pressure per beat, and average nutrient supply at the cellular level—can experience important alterations.

Despite the fact that hemodynamic changes are important and diversified the observed variations in the steady-state 0G spaceflight configuration are not dramatic per se. For instance, although $EF$ reduction was not negligible ($-10\%$), myocardial contractility reduction is considered clinically relevant when $EF$ drops 20% or more, which has not been observed to date during any spaceflight[53]. After long-term spaceflight a 0G adaptation point imposed by a less demanding environment is indeed reached by the cardiovascular system. The scenario becomes potentially hazardous at the time of reentry on the Earth or partial gravity restoration (e.g., Moon/Mars landing). Leaving the 0G environment imparts a backward blood shift from upper-to-lower body. Recalling that mean pressure at central, cerebral, and organ levels was decreased by 10%, backward blood shift combined with an overall hypovolemic condition ($-11.5\%$ for the total blood volume) can impair normal perfusion above VIP. Moreover, compliance increase and resistance decrease in the lower limbs amplify the venous pooling already present during reentry and landing. All these mechanisms converge to produce the well-known condition of orthostatic intolerance, which is further aggravated by the reduced and slowed baroreceptor response. In fact, as baroreceptors are not able to promptly control the

    Published in cooperation with the Biodesign Institute at Arizona State University, with the support of NASA

**Table 4.** *NSD* values at different significant sites for pressure and flow rate signals.

| Cardiovascular region | NSD pressure | NSD flow rate |
|---|---|---|
| Arterial tree pathway (1D) | | |
|   Ascending aorta | 0.19 | 0.28 |
|   Thoracic aorta | 0.24 | 0.37 |
|   Abdominal aorta A | 0.31 | 0.37 |
|   Abdominal aorta E | 0.33 | 0.47 |
|   External iliac | 0.34 | 0.49 |
|   Femoral | 0.37 | 0.56 |
| Cerebral region (0D–1D) | | |
|   HA veins | 0.56 | 0.56 |
|   Left vertebral | 0.25 | 0.32 |
|   Left vertebral arterioles | 0.25 | 0.23 |
|   Left internal carotid | 0.30 | 0.43 |
|   Left internal carotid arterioles | 0.26 | 0.27 |
|   Left external carotid | 0.30 | 0.41 |
|   Left external carotid arterioles | 0.23 | 0.25 |
| Cardio-thoracic region (0D–1D) | | |
|   LA and MV | 0.26 | 0.34 |
|   LV and AV | 0.13 | 0.28 |
|   RA and TV | 0.41 | 0.44 |
|   RV and PV | 0.15 | 0.23 |
|   Pulmonary arteries | 0.16 | 0.17 |
|   Pulmonary veins | 0.31 | 0.23 |
|   SVC | 0.51 | 0.49 |
|   IVC | 0.46 | 0.44 |
|   Intercostals | 0.29 | 0.49 |
|   Intercostals arterioles | 0.26 | 0.28 |
|   Left brachial | 0.29 | 0.42 |
| Abdominal region and organs (0D–1D) | | |
|   Gastric | 0.31 | 0.32 |
|   Gastric arterioles | 0.28 | 0.31 |
|   Celiac A | 0.30 | 0.39 |
|   Left renal | 0.27 | 0.32 |
|   U left renal arterioles | 0.30 | 0.33 |
|   L left renal arterioles | 0.30 | 0.30 |
|   U ABD venules | 0.30 | 0.30 |
|   L ABD venules | 0.33 | 0.33 |
| Lower limbs region (0D–1D) | | |
|   Inner iliac | 0.34 | 0.48 |
|   Inner iliac arterioles | 0.34 | 0.35 |
|   Deep femoral | 0.37 | 0.48 |
|   Deep femoral arterioles | 0.38 | 0.37 |
|   Anterior tibial | 0.48 | 0.54 |
|   Anterior tibial arterioles | 0.45 | 0.60 |
|   Posterior tibial | 0.48 | 0.63 |
|   Posterior tibial arterioles | 0.48 | 0.52 |
|   Legs capillaries | 0.40 | 0.40 |
|   Legs veins | 0.50 | 0.50 |

vasculature and regulate chronotropic and inotropic effects, a normotensive state is hardly guaranteed[3].

Exercise capability was also markedly reduced. A hypothetical Moon or Mars landing (with partial gravity restored) requires a high work capacity and exercise tolerance, which were instead damped with respect to preflight conditions. The heart got used to a reduced pumping requirement and reduced work, with subsequent need for less oxygen. The less demanding spaceflight environment caused a central *PP* reduction of 22.30%. In 1G conditions *PP* is lower for sedentary than resistance-trained people[54–57], thus *PP* can be taken as a marker of training level and physical activity. Gravity is itself a natural source of loading for the human body: the lower spine, lower extremities, and postural muscles are loaded up to 4–5 times the body weight simply walking on the Earth[2]. Therefore, in 1G, strength training is naturally provided by the gravity vector on the body. During spaceflight this natural load is absent, and without any particular countermeasure program, the exercise tolerance of a spaceflight traveler becomes comparable to that of an untrained person with a sedentary lifestyle.

All these aspects, along with other spaceflight-induced effects such as muscular atrophy and bone demineralization, need to be properly addressed especially in long-term human space missions where prompt physical capability after partial or complete gravity restoration is required. The outcomes of the present study highlighted a wealth of cardiovascular information otherwise not yet available in vivo, showing how computational hemodynamics is particularly valuable in environmental conditions such as microgravity and spaceflight where even basic clinical measures are rarely available. It should be also noted that given identification (sex and age) and anthropometric parameters (such as weight and height), together with basic cardiovascular information (such as brachial cuff pressure), the model can be made patient-specific. That would provide good estimates of cardiovascular fitness related to the heterogeneity of the crew, similar to what was already done in 1G physiological and pathological conditions[32,34,58]. Cardiovascular deconditioning without ad hoc countermeasures, as modeled here, is fundamental importance to have a clear and valid baseline for testing against the cardiovascular response with specific countermeasures. By comparing and contrasting the two conditions the efficacy of a single or a combination of countermeasures can be extensively evaluated. A future step might include the effect of one or more countermeasures in the model to observe the response to and help design, implement, and improve such countermeasures. Future modeling developments can also estimate the grade of orthostatic intolerance and exercise tolerance by simulating the return to a gravitational environment, starting from the cardiovascular deconditioned configuration here described. Eventually, understanding the mechanisms of cardiovascular deconditioning during spaceflight will have a significant impact on the knowledge of aging physiology on the Earth[3]. Since deconditioning in spaceflight by gravity deprivation is strongly analogous to deconditioning on the Earth by gravity withdrawal, as in sedentary aging, long-term spaceflight physiology can offer precious hints for delaying or preventing modern lifestyle medical disorders related to increased longevity.

## METHODS

### Mathematical model

The present multiscale modeling of the cardiovascular system combines a 1D description of the arterial tree together with a lumped parameterization of the remaining regions, that is, venous return, heart chambers, pulmonary circulation and baroreceptor regulation (see Fig. 5). The model has been tested using heart pacing and open-loop response, resulting in good agreement with measured data[31]. The hemodynamics of large-to-medium sized arteries relies on the model first proposed and validated by Guala et al.[32], which has been employed to study aging[24,25] and atrial fibrillation effects[28,29].

Mass and momentum balance equations were expressed in one-dimensional form, considering dissipation and inertial terms, and were combined with a constitutive equation accounting for the nonlinear

Published in cooperation with the Biodesign Institute at Arizona State University, with the support of NASA

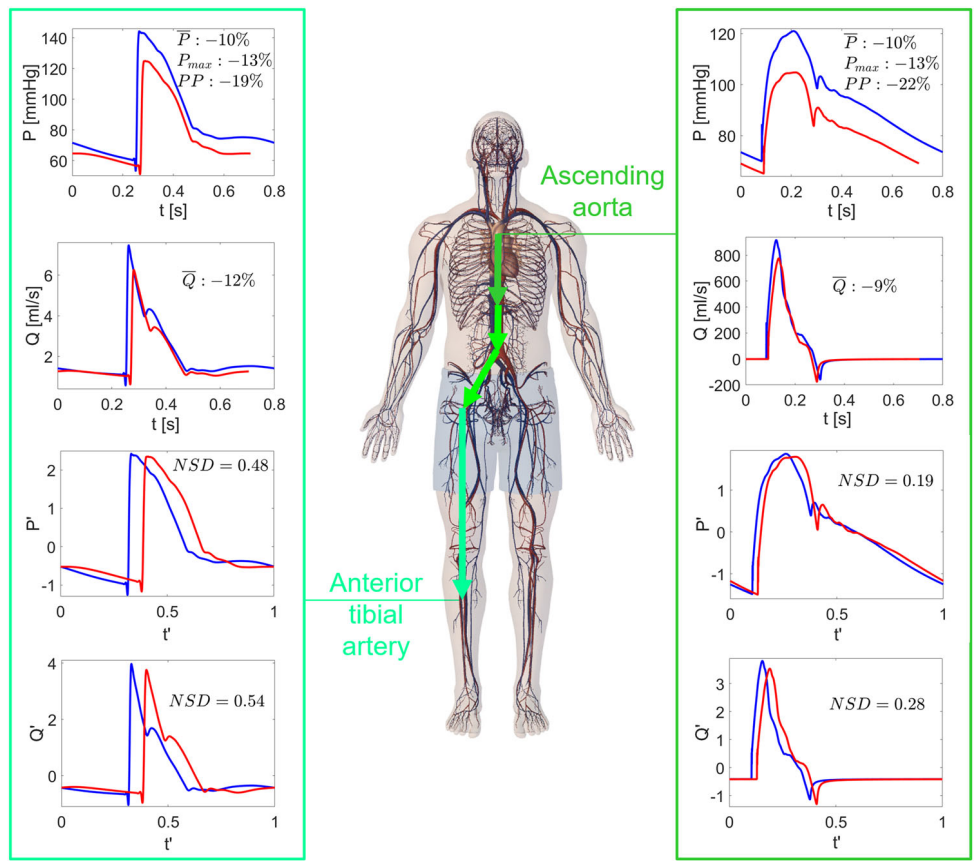

**Fig. 4 Proximal-to-distal arterial tree: pressures and flow rates in 1G supine and 0G conditions.** Time-series $P(t)$, $Q(t)$, $P'(t')$, and $Q'(t')$ at the extremes of the 1D arterial proximal-to-distal tree: ascending aorta and anterior tibial artery. Blue: supine 1G configuration on the Earth; red: 0G spaceflight configuration. *NSD* values for $Q'$ and $P'$, together with mean pressure ($\bar{P}$), mean flow rate ($\bar{Q}$), max pressure ($P_{max}$), and pulse pressure ($PP$) relative variations between 1G and 0G configurations are reported in the panels.

viscoelastic behavior of the tapered arterial walls. Arterioles, capillaries, venous return (venules, veins, and venae cavae), cardiac, and pulmonary (arterial and venous) circulations were modeled through *RLC* 0D compartments, each characterizing local hemodynamics. The *RLC* network accounts for viscous (resistance, *R*), inertial (inductance, *L*), and distensibility/contractility (compliance, *C*, or elastance, *E*) effects. All four cardiac chambers were contractile, as modeled through time-varying elastance functions, including the atrial kick feature. Valvular dynamics were well captured, dominated by the main forces acting on the valves, such as pressure difference across the valve, frictional effects from neighboring tissue resistance, the dynamic motion effect of the blood acting on the valve leaflet and the action of the vortex downstream of the valve. The 1D → 0D interface was at the arteriolar level, with a characteristic impedance per artery put before the corresponding arteriole. The 0D → 1D interface was at the aortic valve level. The position of the volume indifference point (VIP, see the Supplementary Information for more details) split the model into upper and lower body regions (see Fig. 5). The short-term baroregulation mechanisms were also included in the model. The chronotropic and inotropic effects of both ventricles, as well as the control of the systemic vasculature (peripheral arterial resistances, unstressed volumes of the venous system, and venous compliances), were taken into account.

Simulations were run until the steady-state solution was achieved: for both 1G and 0G conditions the transient dynamics were completely extinguished after 100 cardiac periods. Thus, results are referred to the generic steady-state *RR* beat. Details of the governing equations and model parameters are offered in the Supplementary Information.

Definition and nomenclature for the common cardiac parameters are provided in the Supplementary Information. We defined a new metric in the present study, *NSD*. For the waveform analysis of signals, we exploited the dimensionless time $t' = t/RR$ ($t' \in [0, 1]$), where *RR* [s] is the cardiac beating period. We normalized the time-series as $y'(t') = (y(t') - \mu_y)/\sigma_y$, where $\mu$ and $\sigma$ are the mean and standard deviation values of $y(t)$ over the cardiac beat, respectively ($y(t) = P(t)$ or $Q(t)$). We then defined the *NSD* averaged per heartbeat as

$$\mathrm{NSD} = \int_0^1 |y'(t')_{1G} - y'(t')_{0G}|dt', \quad \mathrm{NSD} \in [0, 2],$$

where $y'(t')$ can be either $Q'(t')$ or $P'(t')$ (having zero mean and unitary standard deviation values), and the subscripts 1G and 0G refer to the supine 1G and 0G spaceflight configurations, respectively. As hemodynamic time-series $P'(t')$ and $Q'(t')$ were normalized in vertical (amplitude) and horizontal (time) directions, signals in 1G and 0G can be fully compared: *NSD* quantitatively accounts for the shape and net waveform variation between them. The lower bound value (*NSD* = 0) is reached when the two signals overlap. Thus, if *NSD* is close to 0, no appreciable waveform variation is detected. On the contrary, the upper bound value (*NSD* = 2) can be reached only if the two signals are step functions (centered in $t' = 0.5$ and $y' = 0$ and with half-width equal to 1), one complementary to the other. The closer *NSD* is to 2, the more significant is the alteration of the waveform.

## Spaceflight setting: criteria, cardiovascular mechanisms, and adopted changes

The setting reproducing simulated long-term (at least 5 months) spaceflight conditions without ad hoc countermeasures relies on an extensive bibliographic investigation of more than 50 studies (listed in the Supplementary Information) of human cardiovascular changes induced by microgravity. A general overview of the adopted hemodynamic 0G setting is sketched, while a detailed description is offered in the Supplementary Information. Among hemodynamic data available in literature we focused on those measured during spaceflights rather than ground-based experiments. When no useful information during actual spaceflight was available—such as for blood shift—we considered parabolic flight data, which is more informative since blood shift is extinguished in few hours. Within the subset of spaceflight data priority has been given to long-term and recent space missions. By analyzing the selected literature we based the model setting and validation primarily on long-term data having routine and involuntary countermeasures, as well as countermeasures not expressly declared. We did not directly use (for setting or validation) studies adopting particular ad hoc countermeasures, but accounted for them only in order to better contextualize and qualitatively compare our results. Despite the great caution in collecting and managing spaceflight information, it should be kept in mind that: (i) spaceflight literature data often reports variations with respect to the 1G upright condition, while the present model has been derived in 1G supine condition for clinical interest. For this reason, when possible, further estimates were done to convert 1G upright condition measures to 1G supine conditions; (ii) older data are usually without any countermeasure program while the more recent ones adopt a wide variety of different countermeasures. Although more recent studies were preferred, we took advantage of older measurements (usually without countermeasures) to properly weight the possible effects of countermeasures; (iii) there are overall few long-term missions and data are not always in agreement (usually because of a different type of experimental setting and reference position). This last aspect made the spaceflight setting even more challenging, since a significant part of the analyzed data has been considered not sufficiently reliable for present purposes.

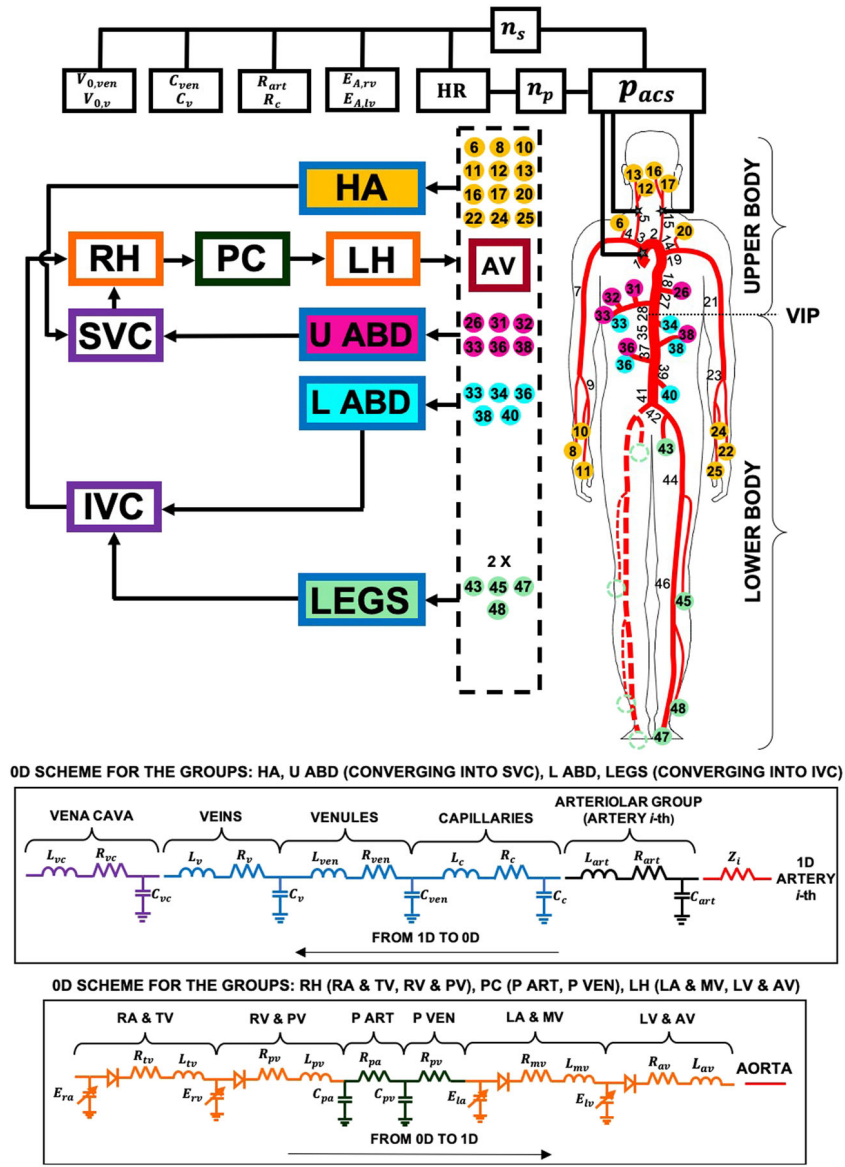

**Fig. 5   Scheme of the multiscale model.** The sketch includes the arterial tree (right), the structure of the 0D compartments (left), the arteriolar 0D–1D interface (dashed box), and the baroreceptor mechanisms (top). Details of the 0D compartments are depicted in the bottom panels. According to VIP, the model is divided into upper body (RH right heart, PC pulmonary circulation, LH left heart, U ABD and HA upper body arteries and arterioles; upper abdomen, head and arms venous return, SVC superior vena cava,) and lower body (L ABD and LEGS lower body arteries and arterioles; lower abdomen and legs venous return, IVC inferior vena cava). Modeling parameters and variables are defined in the Supplementary Information, see Supplementary Table 1 for the legend.

The main cardiovascular changes due to weightlessness introduced into the model were related to blood shift, blood volume reduction, leg venous compliance, cardiac function, arterial resistance, and baroreflex response. Table 5 summarizes the mechanisms considered, the resulting cardiovascular changes and details, where necessary, about their modeling implementation. For blood shift, once VIP was located, blood volume

Published in cooperation with the Biodesign Institute at Arizona State University, with the support of NASA

**Table 5.** Summary of the macroscopic mechanisms considered, the adopted cardiovascular changes with respect to the 1G supine condition and details of the corresponding modeling implementation.

| Cardiovascular processes | Adopted changes | Modeling implementation |
|---|---|---|
| Blood volume shift | Legs volume: −235 ml | Total volumes $V$ of the zones within each region were varied proportionally to the 1G supine condition. Unstressed volumes $V_0$ were varied to preserve $V_0/V$ as in 1G supine condition |
| | Lower abdomen volume: −234 ml | |
| | Head-arms volume: +133 ml | |
| | Cardiac-thoracic volume: +202 ml | |
| | Upper abdomen volume: +134 ml | |
| Blood volume reduction | −11.5% | Within each region, total volumes $V$ were reduced proportionally to the configuration after blood shift. Unstressed volumes $V_0$ were set to preserve $V_0/V$ as in the configuration after blood shift |
| Cardiac function | Amplitude of ventricular elastances: −27% | Cardiac volume reduction was implemented by reducing unstressed volumes $V_0$ of all cardiac chambers by 90% |
| | Min ventricular elastances: +3% | |
| | Pulmonary artery compliance: +4% | |
| | Pulmonary vein compliance: +5% | |
| | Cardiac volume: −10% | |
| Legs venous compliance | +27% | Model parameters are changed of the same percentage |
| Arterial resistances | Vertebral and carotid arteries: +10% | Model parameters are changed of the same percentage |
| | Lower body: −10% | |
| Baroreflex response | Baseline $HR$: +13% | Model parameters are changed of the same percentage |
| | Baseline average aortic-carotid | |
| | sinus pressure: −10% | |

was shifted from lower to upper body using parabolic flight data and based on both the vessel distensibility and the distance from VIP. Total blood volume was then reduced by 11.5%, in accordance with long-term spaceflight data. To account for the contractility and volume reductions ventricular elastances were modified and cardiac volume was decreased. Leg venous compliance was increased by 27%, while arterial resistances were changed according to the body region (+10% for vertebral and carotid arteries; −10% for the lower body), as observed after long exposure to weightlessness. For baroreflex mechanisms only baseline values were altered as we were interested in long-term spaceflight in steady-state conditions: $HR$ was increased by 13% and average aortic-carotid sinus pressure was decreased by 10%. An extended description of the hemodynamic considerations and the resulting estimated variations for each cardiovascular mechanism is detailed in the Supplementary Information, together with the corresponding bibliographic references supporting the adopted changes.

### Limitations

A limiting aspect of the present model is that it does not directly control interstitial fluid shift, since we can only act directly on blood volume and blood shift. Nevertheless, interstitial fluid shift and other long-term mechanisms—such as muscle atrophy and blood volume reduction—were intrinsically accounted for in the setting of the long-term spaceflight configuration by modifying the total and unstressed volumes of all cardiovascular compartments. In this regard, effects—such as net diuresis and fluid re-equilibration, fluid shift from intravascular to interstitial space, and the possibility of reduced thirst—were not individually accounted for, but overall were included in the blood volume reduction mechanism.

The model currently does not explicitly include a gravity force term since its effects were modeled based on the available microgravity literature. A gravity force term can be added to the model equations and adjusted to appropriately reproduce lunar, Martian, or other gravity conditions. At present this use would be purely predictive but very helpful in planning future interplanetary missions and extraterrestrial human occupancy.

### Reporting summary

Further information on experimental design is available in the Nature Research Reporting Summary linked to this article.

### DATA AVAILABILITY

The datasets generated and analyzed during the current study are available from the corresponding author on reasonable request.

    Published in cooperation with the Biodesign Institute at Arizona State University, with the support of NASA

## ACKNOWLEDGEMENTS

The authors would like to thank Claudio Canuto for the precious help with the numerical implementation of the model, Matteo Anselmino for the fruitful discussion of the results, and Mark Miller for his valuable contributions to the editing of the manuscript.

## AUTHOR CONTRIBUTIONS

## COMPETING INTERESTS

## ADDITIONAL INFORMATION

Published in cooperation with the Biodesign Institute at Arizona State University, with the support of NASA